\RequirePackage{fix-cm}
\documentclass[smallextended]{svjour3}       % onecolumn (second format)
\smartqed  % flush right qed marks, e.g. at end of proof
\usepackage{graphicx}
\usepackage{subcaption}
\usepackage{textcomp}
\usepackage{amsfonts}
\usepackage{amsmath}
\usepackage{gensymb}
\usepackage{enumitem}

\renewcommand{\theenumi}{\alph{enumi}}
\newcommand{\R}{\mathbb R}
\newtheorem{thm}{Theorem}
\newtheorem{lem}[thm]{Lemma}

\begin{document}

\title{Voronoi-based similarity distances between arbitrary crystal lattices%\thanks{Grants or other notes
}


\author{M. M. Mosca         \and
        Dr. V. Kurlin %etc.
}


\institute{M. M. Mosca \and Dr. V. Kurlin \at
              Materials Innovation Factory and Computer Science department,\\
              University of Liverpool, Liverpool L69 3BX, United Kingdom\\
              \email{vitaliy.kurlin@liverpool.ac.uk}           %  \\
}

\date{Received: date / Accepted: date}

\maketitle

\begin{abstract}
This paper develops a new continuous approach to a similarity between periodic lattices of ideal crystals. Quantifying a similarity between crystal structures is needed to substantially speed up the Crystal Structure Prediction, because the prediction of many target properties of crystal structures is computationally slow and is essentially repeated for many nearly identical simulated structures. The proposed distances between arbitrary periodic lattices of crystal structures are invariant under all rigid motions, satisfy the metric axioms and continuity under atomic perturbations. The above properties make these distances ideal tools for clustering and visualizing large datasets of crystal structures. All the conclusions are rigorously proved and justified by experiments on real and simulated crystal structures reported in the Nature 2017 paper “Functional materials discovery using energy–structure–function maps”. 
\keywords{crystal lattices \and similarity distances \and Voronoi cells \and crystal structure prediction}
\end{abstract}

\section{Introduction: motivations for similarity distances from Crystal Structure Prediction}
\label{sec:intro}
Modern tools of the Crystal Structure Prediction (CSP) produce large datasets of thousands or even millions of simulated crystals based on the same chemical composition. \cite{biblio:faraday} Many of these crystal structures are geometrically similar, because they were obtained as approximations to local minima of a complicated energy. The available similarity tests miss too many nearly identical structures that are ambiguously represented in different ways. That is why Prof Sally Price has summarized the state-of-the-art in CSP as `the embarrassment of over-prediction’.\cite{biblio:faraday} \\

A dream CSP solution for the pharma industry would be a reliable method to output a short list of only few most stable polymorphs based on a given molecular input. \cite{biblio:faraday} The drug design will be substantially sped up if one can enrich any dataset of simulated crystal structures by a justified distance that shows which structures are geometric neighbors, i.e. close to each other according to this distance, and how such neighborhoods are located relatively to each other.\\
The astronomical analogy is to consider individual crystal structures as stars that form neighborhoods or galaxies. A distance information between visible stars (crystals) will allow one to visualize the whole universe. Such a geometric map of the crystal space \cite{biblio:linjiangandy} will enable a guided navigation in hot spots to further improve properties of known crystals; or a better search in unexplored regions that can contain exotic materials with extraordinary properties.\\
\\
This paper proposes two new distances between arbitrary crystal lattices that are not restricted to the same crystal system or a Bravais type. All lattices can be continuously deformed into each other. Hence a similarity distance should be well-defined on the whole space of lattices. Section 2 defines necessary concepts and states the equivalence and distance problems for crystals and lattices. Section 3 discusses past approaches to similarities of crystal structures.
Section 4 introduces three distances based on the Voronoi cell of an arbitrary lattice. Section 5 shows experimental results on the T2 dataset of simulated and real crystal structures that consists of nano-porous crystals structures that are all based on a single T2 molecule.\cite{biblio:linjiangandy}

\section{Rigorous definitions and problem statements for crystal equivalences and distances}
\label{sec:defs_probstatement}
This section formalizes concepts of periodic crystals, lattices, equivalences and distances.
\textbf{\subsection{The comparison (equivalence) problem for periodic crystals}}
\label{sec:defs_probstatement:comparison}

The general model of an ideal crystal is a periodic cloud of zero-sized points representing atoms. The periodicity is determined by a lattice whose nodes are abstract points, not atoms.\\
A \textit{lattice} in the Euclidean space $\R^3$ is a set of points given by integer linear combinations $x\mathbf{u}+y\mathbf{v}+z\mathbf{w}$ of 3 basis vectors $\mathbf{u,v,w}$, where $x,y,z$ are integer coefficients. If $x,y,z$ are restricted to the interval [0,1] in the real line, the points $x\mathbf{u}+y\mathbf{v}+z\mathbf{w}$ form a primitive unit cell, which is a parallelepiped or a non-rectangular box with parallel opposite sides. The same lattice can be generated by infinitely many bases, hence can have many different unit cells.\\
A \textit{periodic crystal} is defined by a lattice \textit{L} and a \textit{motif}, which is a collection of molecules (for molecular crystals) or atoms or ions (in the case of a non-molecular crystals such as NaCl). The motif is periodically translated in the directions along the 3 vectors that define a unit cell of \textit{L}. Because of many possible unit cells, it is not immediately obvious to decide if crystal structures (or lattices) represented by different unit cells are equivalent in the sense below.

\textbf{\subsection{The equivalence problem and geometric invariants for periodic crystals and lattices}}
\label{sec:defs_probstatement:eqprob_geoinv}

Crystals are often represented by Crystallographic Information Files (CIFs), which contains edge-lengths and angles of a unit cell \textit{U} and fractional coordinates of atoms in the basis of \textit{U}, i.e. as numbers within the interval [0,1]. These coordinates are often given for an asymmetric unit that generates a full motif in \textit{U} by applying symmetry operations specified in a CIF. \\
Crystals are solid materials, hence are invariant (unaffected) by rigid motions in $\R^3$, which are
compositions of rotations and translations. Hence any comparison of crystals should take into account infinitely many positions (of a crystal or its lattice) related by rigid motions in $\R^3$.\\
Crystals structures (or lattices) are called equivalent (or isometric) $\R^3$ if they can be obtained from each other by a rigid motion, which preserves distances between any points in $\R^3$.\\
This equivalence is the minimal possible one to study crystals as solid materials. For example, if atom positions are perturbed, the perturbed crystal (or its lattice) can be geometrically different even if only slightly. So, the space of equivalence classes of lattices under rigid motions is infinite and continuous (or connected). Hence quantifying a similarity between perturbed crystals is an important problem, which is formalized in the next subsection.\\
The Bravais classification puts lattices into a much smaller number of classes (only 14 types in dimension 3), though lattices from different classes can be geometrically nearly identical.\\
Two randomly chosen lattices will share only the translation group of symmetries, hence we need other tools to check if given lattices are not equivalent. Such classification tools are called invariants. An \textit{invariant} of lattices up to a certain equivalence relation, for example rigid motions, is a function that should take the same value on all equivalent lattices.\\
For example, the volume of a primitive unit cell is an invariant, because all primitive unit cells of a given lattice have the same volume. Edge-lengths and angles of a unit cell are not invariants, because there are infinitely many primitive cells that define the same lattices.\\
The equivalence problem for lattices is to design a robust algorithm that accepts two arbitrary lattices (without any extra parameters) and decides whether they are equivalent or not. Theoretically, such an algorithm can be based on Niggli’s reduced cells  \cite{biblio:niggli} in subsection 3.3, and their instability under perturbations \cite{biblio:andrews_acta} motivates the harder distance problem below.

\textbf{\subsection{The distance problem for lattices of periodic crystal structures}}
\label{sec:defs_probstatement:distanceprob}
This subsection states the metric axioms that are needed to successfully map the space of all crystal structures for any given composition of molecules, atoms or ions. If a distance function between crystal structures satisfies metric axioms below, the crystallography will be open to rigorous methods of metric geometry that will measure what portions of a crystal space are explored and what regions require more sampling in computer simulations. 
Let $\R_+$ denote the set of all non-negative real numbers. Let \textit{S} be any set, e.g. \textit{S} can be any collection of crystal structures or lattices. A distance (or a metric) on \textit{S} is a function $d:S \times S \rightarrow \R_+$, such that the following conditions (or metric axioms) hold:
\begin{enumerate}[label=\textbf{(2.3\alph*)}, leftmargin=\parindent,align=left,labelwidth=\parindent,labelsep=0pt]	
	\item \label{axioms:identity}
	\quad for any $C,C' \in S$, the distance $d(C,C')=0$ if and only if $C,C'$ are equivalent (equal);
	
	\item \label{axioms:symmetry}
	\quad symmetry: $d(C,C')=d(C',C)$ for any $C,C' \in S$;
	
	\item \label{axioms:triangle_ineq}
	\quad triangle inequality: $d(C,C')+d(C',C'') \geq d(C,C'')$ for any $C,C',C'' \in S$.
\end{enumerate}
For $S=\R^n$, one will use the Euclidean space $$d(p,q)=\sqrt{(p_1-q_1 )^2+...+(p_n-q_n )^2}$$ between points $p=(p_1,...,p_n)$ and $q=(q_1,...,q_n)$, which satisfies the axioms above. \\

For a set \textit{S} of crystal structures or arbitrary lattices, it is a hard problem to define a distance function \textit{d} satisfying the axioms above, because \textit{d} should not depend on a way to represent crystal structures or lattices, hence should be independent of many potential unit cells. \\
Axiom \ref{axioms:identity} avoids trivial examples when a distance \textit{d} is constant, i.e. has the same value $d(C,C')$ for any non-equivalent crystal structure $C \neq C'$. Axiom \ref{axioms:symmetry} says that a distance remains the same if endpoints are swapped. Axiom \ref{axioms:triangle_ineq} is motivated by the assumption that a shortest path from \textit{C} to \textit{C''} should not be longer than a combination of shortest paths from \textit{C} to \textit{C'} and then from \textit{C'} to \textit{C''}. The metric axioms are claimed to be checked for the Euclidean distance between fingerprints proposed by Zhu et al. \cite{biblio:zhu_amsler} Any such approach should justify that any non-equivalent crystal structures \textit{C,C'} have different feature vectors. Else the distance between identical vectors of non-equivalent crystal structures \textit{C,C'} is 0 and axiom \ref{axioms:identity} fails.\\
\\
The \textit{distance problem} for crystal structures (or their lattices) is to find a distance function that satisfies metric axioms \ref{axioms:identity}, \ref{axioms:symmetry}, \ref{axioms:triangle_ineq} above and also the continuity condition below:
\begin{enumerate}[label=\textbf{(2.3\alph*)}, leftmargin=\parindent,align=left,labelwidth=\parindent,labelsep=0pt]
	\setcounter{enumi}{3}	
	\item \label{axioms:continuity}
	\quad the distance $d(C,C')$ continuously changes under perturbations of crystal structures, e.g. if cell parameters or atomic positions are noisy; in particular, the range of \textit{d} should be a continuous interval, possibly $[0,+\infty)$, but not only a finite collection of discrete values.
\end{enumerate}
One more potentially useful property of similarity distance is invariance under scaling below:
\begin{enumerate}[label=\textbf{(2.3\alph*)}, leftmargin=\parindent,align=left,labelwidth=\parindent,labelsep=0pt]
	\setcounter{enumi}{4}	
	\item \label{axioms:scaling} 
	\quad the distance $d(C,C')$ should remain unchanged if both sets $C,C' \subset R^n$ are scaled by the same factor $s>0$, i.e. $d(C,C')=d(sC,sC')$, where $sC=\{ s\:p \in \R^n: \textrm{for any point } p \in C\}$.
\end{enumerate}
\bigskip

\section{Strengths and weaknesses of past approaches to a similarity of crystal structures}
\label{sec:approaches}
\textbf{\subsection{The COMPACK algorithm for the Cambridge Structural Database (CSD)}}
\label{sec:approaches:compaq}
The widely used COMPACK algorithm \cite{biblio:chrisholm_j_appl_cryst} for identifying crystal structure similarity requires specified tolerances, e.g. 15\% on distance constraints, relative to a reference structure \textit{S}, and outputs a list of other structures (from a given dataset such as CSD) that are found to be close to the reference \textit{S}. The COMPACK output is a single set of similar crystal structures, though a continuous hierarchy ordered by distances to the reference \textit{S} would be more informative.\\
A numerical measure of similarity for two crystal structures offered by the Mercury software equals the root mean square deviation of atomic positions over finitely many (up to 15 by default) matched molecules or atoms. If this partial matching is extended to the full infinite crystal structures, the deviation of positions will infinitely grow, hence is defined only for finite portions, not for equivalence classes of periodic lattices considered up to rigid motions.

\textbf{\subsection{The COMPSTRU tool at the Bilbao Crystallographic Server (BCS)}}
\label{sec:approaches:compstru}

Similarly to COMPACK, the recent COMPSTRU algorithm \cite{biblio:flor_j_appl_cryst} measures a similarity between a given reference structure \textit{S} and crystal structures whose lattice parameters should be close to those of \textit{S} (by default 0.5A for distances and 5\textdegree for angles). This comparison is restricted to crystal structures that have the same space-group type. A slight perturbation of atomic positions of the reference \textit{S} will produce a nearly identical crystal that is not comparable to the reference \textit{S}, hence continuity condition \ref{axioms:continuity} is not satisfied. Many other approaches are based on closest parameters of crystal structures or reduced unit cells discussed below.

\textbf{\subsection{Comparison algorithms based on reduced cells of crystal lattices}}
\label{sec:approaches:reducedcell}

\begin{figure}[h]
	\centering%
	\begin{subfigure}[b]{0.25\textwidth}
		\centering% <-- added
		\includegraphics[width=\textwidth, keepaspectratio]{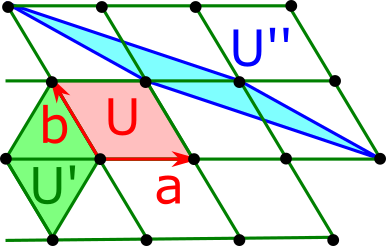}
		\label{fig:cells}
	\end{subfigure}%
	\begin{subfigure}[b]{0.25\textwidth}
		\centering % <-- added
		\includegraphics[width=\textwidth, keepaspectratio]{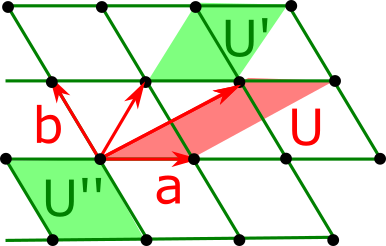}
		\label{fig:niggli_reduction}
	\end{subfigure}
	\begin{subfigure}[b]{0.25\textwidth}
		\centering % <-- added
		\includegraphics[width=\textwidth, keepaspectratio]{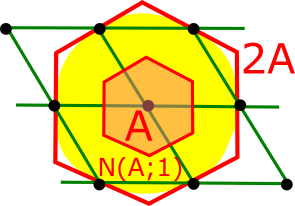}
		\label{fig:hex_voronoi_offsets}
	\end{subfigure}
	\begin{subfigure}[b]{0.2\textwidth}
		\centering % <-- added
		\includegraphics[width=\textwidth, keepaspectratio]{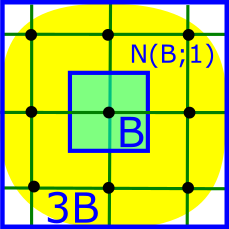}
		\label{fig:square_voronoi_offsets}
	\end{subfigure}
	%\captionsetup{justification=centering}
	\caption{(1st: any lattice has infinitely many primitive cells, e.g. \textit{U,U',U''}. 2nd: Niggli's reduction of a vector $v_2$ relative to $v_1$ can lead to two cells \textit{U'} and \textit{U''}, a choice is unstable. 3rd: the yellow offset \textit{N(A;1)} of A is inside the scaled cell 2A. 4th: yellow offset \textit{N(B;1)}.)}
	\label{fig:cells_voronoioffsets}	
\end{figure}

Despite any lattice can be defined by infinitely many primitive unit cells, Niggli introduced a reduced cell, which is unique and can be theoretically used for comparing lattices. \cite{biblio:niggli} Niggli’s reduced cell is unstable under perturbations in the sense that a reduced cell of a perturbed lattice can have a basis that substantially differs from that of a non-perturbed lattice. \cite{biblio:andrews_acta}\\
\textbf{Figure \ref{fig:cells_voronoioffsets}} illustrates the 2-dimensional case of Niggli’s reduction, where a narrow an original narrow unit cell is reduced by subtracting multiples of a horizontal basis vector \textbf{a} from another non-horizontal basis vector \textbf{b} until the projection of \textbf{b} to a is close to 0, i.e. fits the interval $[-0.5|\textbf{a}|,0.5|\textbf{a}|]$. The endpoints $\pm0.5|\textbf{a}|$ correspond to two equivalent choices of \textbf{b}. Excluding one of these endpoints will give a unique basis \textbf{\textit{a,b}}, but will make the reduction discontinuous (unstable), because the two nearly identical vectors \textbf{b} whose projections are equal to $0.5|\textbf{a}|\pm \delta$ for any tiny $\delta>0$ will be reduced to different vectors, hence fractional coordinates of atoms in nearly identical crystal structures will have very different values. Many software tools offer parameters that shift the perturbation problem to other bounds of these parameters. The underlying reason of instability is similar to the choice of a range for angles that can be measured within [0,360\textdegree) or within (-180\textdegree,180\textdegree]. The distance between angles should be measured along a shortest round arc on a unit circle, not by breaking a circle into an interval and taking the difference of angle values from this interval. Indeed, for any choice of an interval, some angles that are close within a circle become distant in the interval.\\
Finally, crystal structures are compared (not successfully for Heusler structures \cite{biblio:oliynyk_chem_of_mat}) by powder diffraction patterns up to a cut off radius, which brings discontinuities similarly to other parameters. To the best of our knowledge there was no distance on equivalence classes of crystal structures that satisfies all metric axioms \ref{axioms:identity}, \ref{axioms:symmetry}, \ref{axioms:triangle_ineq} and continuity \ref{axioms:continuity}.

\section{New continuous Voronoi-based distances between arbitrary crystal lattices}
\label{sec:newdistances}
This section starts from reminding auxiliary notions of the Hausdorff distance and Voronoi cells that are needed to introduce two new distances on crystal lattices in subsection 4.3.\\
\textbf{\subsection{Offsets and the Hausdorff distance between arbitrary sets and crystal structures}}
\label{sec:newdistances:offsets_hausdorff}
For any subset $C \subset \R^n$, its \textit{r-offset N(C;r)} consists of all points $p \in \R^n$ that are at a Euclidean distance at most \textit{r} from \textit{C}, i.e. $$N(C;r)=\{p \in \R^n\:|\: d(p,q) \leq r \: for \: some \: q \in C\}$$ If \textit{C} is one point, \textit{N(C;r)} is the ball with the center at \textit{C} and radius r. If \textit{C} is a set of points, \textit{N(C;r)} is the union of balls that have the radius r and centers at all points of \textit{C}. A crystal consisting of the same atoms can be visualized as \textit{N(C;r)}, where r is a bond atomic radius.\\
The Hausdorff distance $d_H$ between any subsets $C,C' \subset \R^n$ is the minimum $r \geq 0$ such that
r-offsets of \textit{C,C'} cover each other, i.e. $C' \subset N(C;r)$ and $C \subset N(C';r)$, or \cite{biblio:kurlin_analysis_images}
\begin{equation}
\label{eq:hausdorff}
d_H(C,C')=\min\{ r \geq 0: C' \subset N(C;r)\:and\:C \subset N(C';r)\}
\end{equation}
For example, $d_H(C,C')=0$ means that $C \subset C'$ and $C' \subset C$, hence $C=C'$ as needed in axiom  \ref{axioms:identity}. This Hausdorff distance was previously used for comparing feature vectors of crystals \cite{biblio:collins_cryst_eng} rather than for crystals. The metric axioms will be proved in the appendix for the extended Hausdorff distance between equivalence classes of lattices introduced in subsection \ref{sec:newdistances:ext_hausdorff}.

\textbf{\subsection{The Voronoi cell of an arbitrary crystal lattice and its geometric stability}}
\label{sec:newdistances:vcell_stability}

\begin{figure}[h]
	\centering%
	\begin{subfigure}[b]{0.13\textwidth}
		\centering% <-- added
		\includegraphics[width=\textwidth, keepaspectratio]{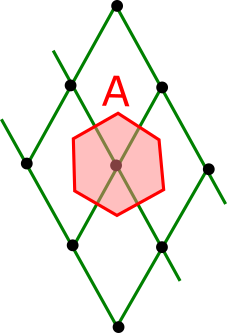}
		\label{fig:hexagonal_latt_vor}
	\end{subfigure}%
	\begin{subfigure}[b]{0.2\textwidth}
		\centering % <-- added
		\includegraphics[width=\textwidth, keepaspectratio]{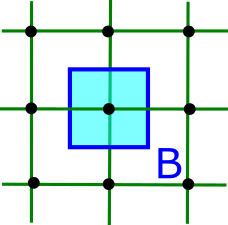}
		\label{fig:square_latt_vor}
	\end{subfigure}
	\begin{subfigure}[b]{0.2\textwidth}
		\centering % <-- added
		\includegraphics[width=\textwidth, keepaspectratio]{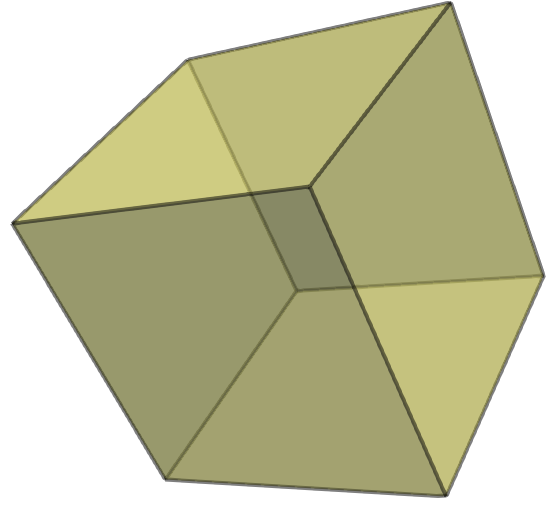}
		\label{fig:cubic_latt_vor}
	\end{subfigure}
	\begin{subfigure}[b]{0.2\textwidth}
		\centering % <-- added
		\includegraphics[width=\textwidth, keepaspectratio]{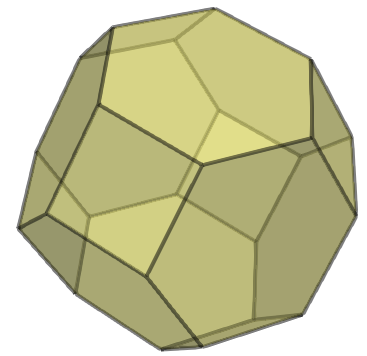}
		\label{fig:bcc_latt_vor}
	\end{subfigure}
	\begin{subfigure}[b]{0.2\textwidth}
		\centering % <-- added
		\includegraphics[width=\textwidth, keepaspectratio]{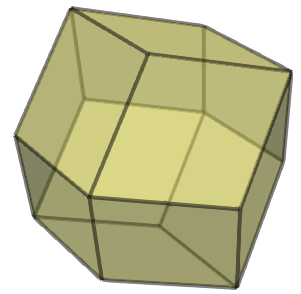}
		\label{fig:fcc_latt_vor}
	\end{subfigure}
	%\captionsetup{justification=centering}
	\caption{(The Voronoi cells of lattices: hexagonal, square, cubic, body-centered cubic (BCC), face-centered cubic (FCC), see definitions in subsection 4.2. the cubic lattice has the standard basis (1,0,0),(0,1,0),(0,0,1); the body-centered cubic (BCC) lattice has the basis (1,0,0),(0,1,0),(0.5,0.5,0.5); the FCC lattice has the basis (1,0,0),(0.5,0.5,0),(0.5,0,0.5). )}
	\label{fig:hex_sq_cub_bcc_fcc_vor}	
\end{figure}

Fix an origin 0 in a lattice $L \subset \R^n$. The \textit{Voronoi cell V(L)} is the set of all points $p \in \R^n$ that are (not strictly) closer to 0 (in the usual Euclidean distance \textit{d}) than to all other points of \textit{L}, i.e. (see \textbf{Figure \ref{fig:hex_sq_cub_bcc_fcc_vor}})
\begin{equation}
\label{eq:def_voronoi}
V(L)=\{p \in \R^n |\:d(p,0) \leq d(p,q)\:for\:any\:q \in L-0\}
\end{equation}
For a 2D lattice, the Voronoi cell is a rectangle or a centrally symmetric hexagon. \cite{biblio:edelsbrunner_comp_geom} For a 3D lattice, a generic Voronoi cell is a truncated octahedron with 7 pairs of parallel faces. The Voronoi cell is centrally symmetric since any lattice is symmetric with respect to its origin. The combinatorial type of \textit{V(L)} can change under perturbations. But the geometric shape of \textit{V(L)} changes continuously by Reem’s theorem below. A similar stability for Voronoi-based skeletons is known for point clouds. \cite{biblio:kurlin_analysis_images} The Voronoi cell of a lattice can be computed from a finite set $L_{[3]}$ of lattice nodes within a factor 3 extension of a suitably reduced \cite{biblio:dolbilin_math_hungarica} cell of \textit{L}. \\
Geometric stability of Voronoi cells (simplified Reem’s Theorem 5.1 \cite{biblio:reem_27symp_geometry}) For any small $\epsilon>0$, there is $r>0$ such that any lattices \textit{L,L'} with a small Hausdorff distance $d_H(L_{[3]} ,L_{[3]}')<r$ should have close Voronoi cells with a small Hausdorff distance $d_H(V(L),V(L'))<\epsilon$.

\textbf{\subsection{The rotational extension of the Hausdorff distance to equivalence classes of lattices}}
\label{sec:newdistances:ext_hausdorff}

\begin{figure}[h]
	\centering%
	\begin{subfigure}[b]{0.16\textwidth}
		\centering% <-- added
		\includegraphics[width=\textwidth, keepaspectratio]{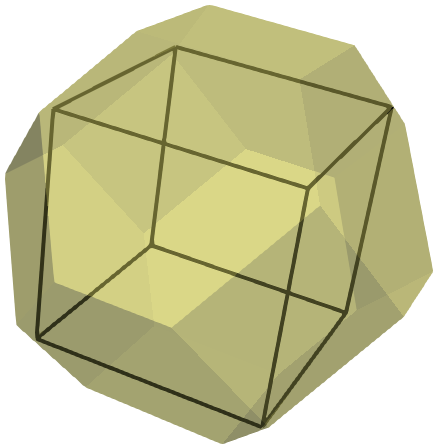}
		\label{fig:ext_hausdorff:cubic_bcc}
	\end{subfigure}%
	\begin{subfigure}[b]{0.16\textwidth}
		\centering % <-- added
		\includegraphics[width=\textwidth, keepaspectratio]{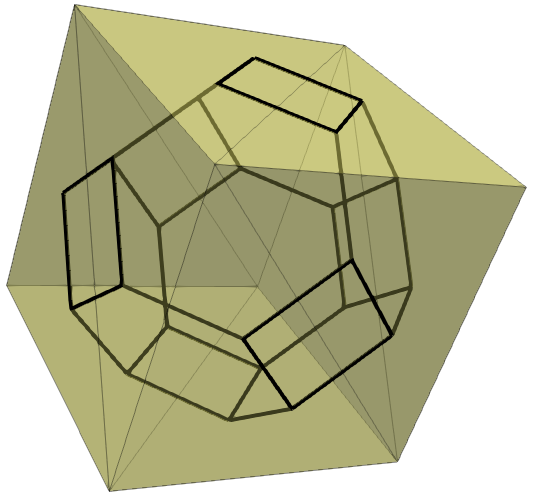}
		\label{fig:ext_hausdorff:bcc_cubic}
	\end{subfigure}
	\begin{subfigure}[b]{0.16\textwidth}
		\centering % <-- added
		\includegraphics[width=\textwidth, keepaspectratio]{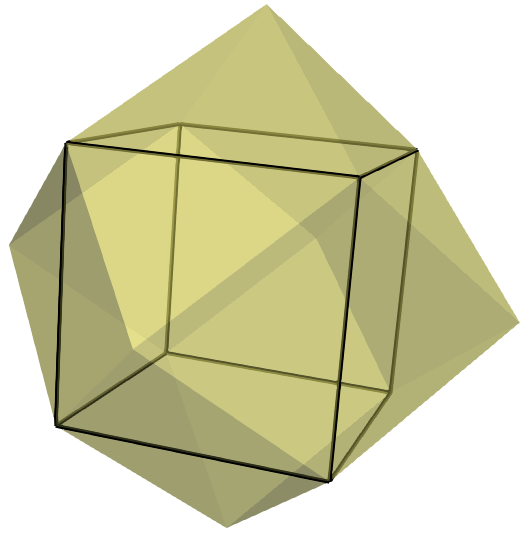}
		\label{fig:ext_hausdorff:cubic_fcc}
	\end{subfigure}
	\begin{subfigure}[b]{0.16\textwidth}
		\centering % <-- added
		\includegraphics[width=\textwidth, keepaspectratio]{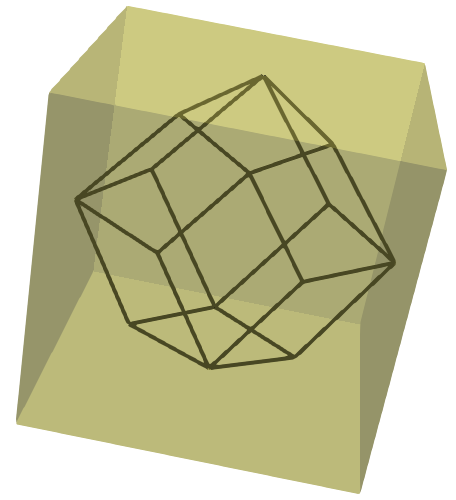}
		\label{fig:ext_hausdorff:fcc_cubic}
	\end{subfigure}
	\begin{subfigure}[b]{0.16\textwidth}
		\centering % <-- added
		\includegraphics[width=\textwidth, keepaspectratio]{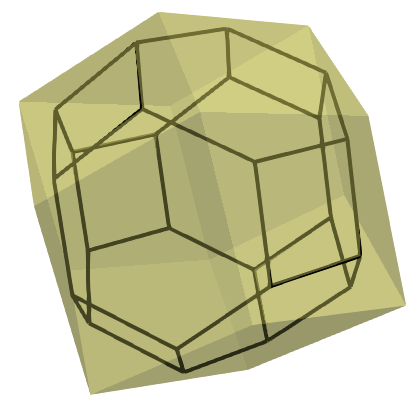}
		\label{fig:ext_hausdorff:bcc_fcc}
	\end{subfigure}
	\begin{subfigure}[b]{0.16\textwidth}
		\centering % <-- added
		\includegraphics[width=\textwidth, keepaspectratio]{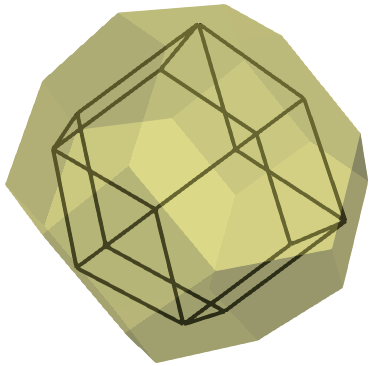}
		\label{fig:ext_hausdorff:fcc_bcc}
	\end{subfigure}
	%\captionsetup{justification=centering}
	\caption{(Illustrations of extended Hausdorff distances: one Voronoi cell \textit{V(L)} is optimally rotated and inscribed into a minimal offset Voronoi cell \textit{V(L')} for pairs \textit{L,L'} from the left to right: (cubic, BCC), (BCC, cubic), (cubic, FCC), (FCC, cubic), (BCC, FCC), (FCC, BCC). )}
	\label{fig:ext_hausdorff:vor_cells_distances}	
\end{figure}

The geometric stability above holds for fixed lattices without equivalences up to rigid motions of $\R^3$. For example, if a lattice \textit{L} is shifted or rotated to a new position, then the new lattice \textit{L'} remains equivalent to \textit{L}, but the Hausdorff distance between non-identical Voronoi cells is positive:\\ $d_H(V(L),V(L'))>0$, which contradicts axiom \ref{axioms:identity} of a metric. 
The key idea of the proposed extension from fixed lattices to their equivalence classes (up to rigid motions) is to minimize the distance over all possible rigid motions. Geometrically, the standard Hausdorff distance measures how much one should enlarge each cell to fit into another. The extended distance $d_H$ finds a minimal enlargement to fit one Voronoi cell into another over all rigid motions. Lemma \ref{lem:ext_hausdorff:centr_sym_polyhedron_and_offset_param} shows that a translation minimizing the Hausdorff distance makes the centers of Voronoi cells identical, which justifies the definition of the extended Hausdorff distance $d_H$ as the minimum over only rotations about the origin, without any translations.\\
\begin{lem}
\label{lem:ext_hausdorff:centr_sym_polyhedron_and_offset_param}
(proved in Appendix). For any centrally symmetric polyhedra\\ $P,P' \subset \R^n$ and a translation $T_v$ by a vector $v \in \R^n$, the offset parameter $$min\{ r: T_v(P) \subset N(P';r)\}$$ is minimal when $T_v$ moves the center $c(P)$ of the polyhedron $P$ to the center $c(P')$ of $P'$.
\end{lem}
By Lemma \ref{lem:ext_hausdorff:centr_sym_polyhedron_and_offset_param} the extended Hausdorff distance can be minimized only over all rotations around the common center (say, the origin in $\R^n$) of the Voronoi cells $V(L),V(L')$ as follows. Rigid motions that preserve the origin in $\R^n$ are defined by special orthogonal $n \times n$ matrices A such that the determinant of A is 1 and the inverse matrix $A^{-1}$ equals the transpose matrix $A^T$. All these matrices form the group denoted by \textit{SO(n)}. The group \textit{SO(3)} consists of rotations around axes that passes through the origin in $R^3$. Define the non-symmetric offset
\begin{equation}
\label{eq:ext_hausdorff:offset}
offset(L,L')=\min \{ r \geq 0 :\:R(V(L)) \subset N(V(L');r) \}
\end{equation}
where the minimum is taken over all rotations  $R \in SO(n)$. The Extended Hausdorff distance for lattices is defined as the symmetric maximum of the two offset parameters, see \textbf{Figure \ref{fig:ext_hausdorff:vor_cells_distances}}: 
\begin{equation}
\label{eq:ext_hausdorff:dH}
d_H(L,L')=\max\{ offset(L,L'),offset(L',L) \}
\end{equation}
\begin{thm}
\label{thm:ext_hausdorff:satisfying_axioms}
(proved in Appendix). The extended Hausdorff distance $d_H$ is independent of a lattice representation and satisfies the axioms \ref{axioms:identity}, \ref{axioms:symmetry}, \ref{axioms:triangle_ineq} and condition \ref{axioms:continuity}.
\end{thm}

\textbf{\subsection{The scaling distance and its rotational extension to equivalence classes of lattices}}
\label{sec:newdistances:scaleinvariant}

\begin{figure}[h]
	\centering%
	\begin{subfigure}[b]{0.16\textwidth}
		\centering% <-- added
		\includegraphics[width=\textwidth, keepaspectratio]{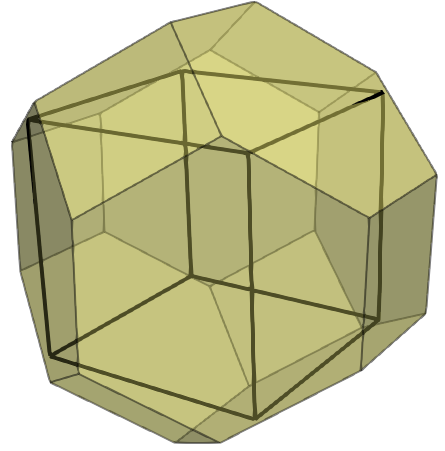}
		\label{fig:scaleinvariant:cubic_bcc}
	\end{subfigure}%
	\begin{subfigure}[b]{0.16\textwidth}
		\centering % <-- added
		\includegraphics[width=\textwidth, keepaspectratio]{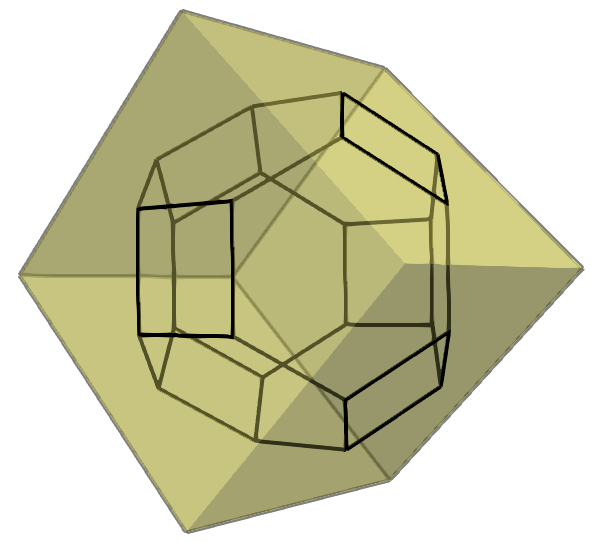}
		\label{fig:scaleinvariant:bcc_cubic}
	\end{subfigure}
	\begin{subfigure}[b]{0.16\textwidth}
		\centering % <-- added
		\includegraphics[width=\textwidth, keepaspectratio]{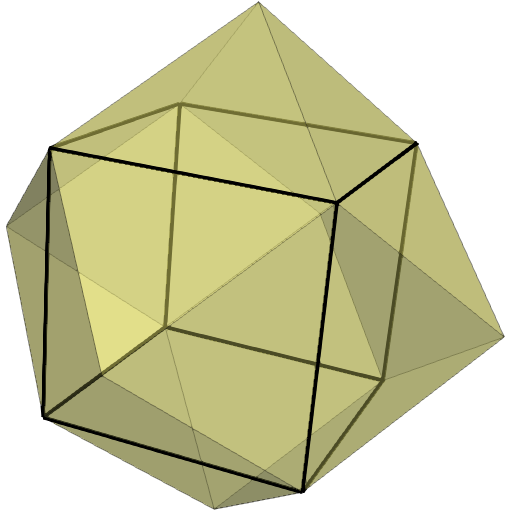}
		\label{fig:scaleinvariant:cubic_fcc}
	\end{subfigure}
	\begin{subfigure}[b]{0.16\textwidth}
		\centering % <-- added
		\includegraphics[width=\textwidth, keepaspectratio]{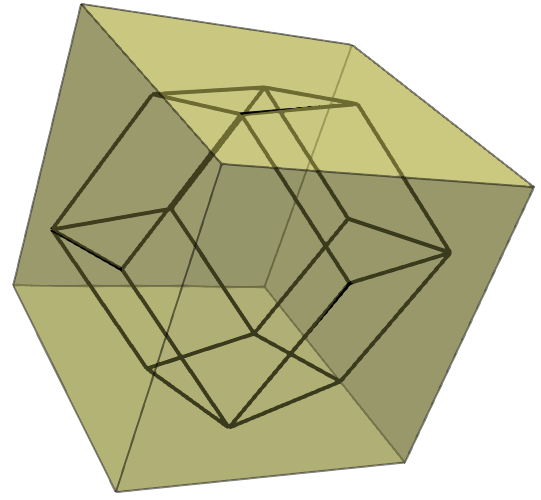}
		\label{fig:scaleinvariant:fcc_cubic}
	\end{subfigure}
	\begin{subfigure}[b]{0.16\textwidth}
		\centering % <-- added
		\includegraphics[width=\textwidth, keepaspectratio]{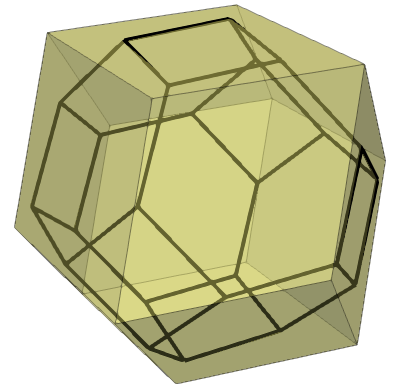}
		\label{fig:scaleinvariant:bcc_fcc}
	\end{subfigure}
	\begin{subfigure}[b]{0.16\textwidth}
		\centering % <-- added
		\includegraphics[width=\textwidth, keepaspectratio]{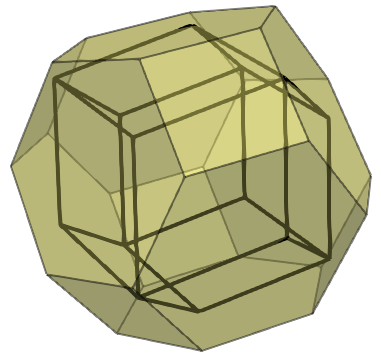}
		\label{fig:scaleinvariant:fcc_bcc}
	\end{subfigure}
	%\captionsetup{justification=centering}
	\caption{(Illustrations of scale-invariant distances $d_s(L,L')$ for pairs \textit{L,L'} from the left to right: (cubic, BCC), (BCC, cubic), (cubic, FCC), (FCC, cubic), (BCC, FCC), (FCC, BCC).)}
	\label{fig:scaleinvariant:vor_cells_distances}
\end{figure}

The Hausdorff distance is additive in the sense that if lattices \textit{L,L'} are scaled by the same factor $s>0$, then $d_H(sL,sL')=s\:d_H(L,L')$. To get a distance invariant under scaling as in condition \ref{axioms:scaling}, the new distance  $d_s$ is introduced first for arbitrary subsets $C,C' \subset \R^n$. Set $$scale(C,C')=min\{ s>0: C \subset sC' \}$$. Then $\ln(\max\{ scale(C,C'),scale(C',C) \})$ satisfies all metric axioms, which is proved below in a more general case for rigid motions.\\
For any lattices \textit{L,L'}, their Voronoi cells are used to minimize two symmetric scales for \textit{V(L),V(L')} over all rigid motions as in subsection \ref{sec:newdistances:ext_hausdorff}. Lemma \ref{lem:scaleinvariant:min_s_centers_superimposed} below similarly to Lemma \ref{lem:ext_hausdorff:centr_sym_polyhedron_and_offset_param} justifies that an optimal translation makes the centers of Voronoi cells identical.\\
\begin{lem}
\label{lem:scaleinvariant:min_s_centers_superimposed}
(proved in Appendix). For any centrally symmetric polyhedra\\ $P,P' \subset \R^n$ and the translation $T_v$ by a vector $v \in \R^n$, the scale factor $$\min\{ s>0: T_v(P) \subset sP'\}$$ is minimal when $T_v$ moves the center \textit{c(P)} of the polyhedron \textit{P} to the center \textit{c(P')} of the polyhedron \textit{P'}.
\end{lem} 
Now the scale factor s can be minimized only over all rotations $R \in SO(n)$ as follows:
\begin{equation}
\label{eq:scaleinvariant:scale}
scale(L,L')=\min\{ s>0: R(V(L)) \subset s\:V(L') \}
\end{equation}
The dimension-less scale-invariant distance between equivalence classes of lattices is (see \textbf{Figure \ref{fig:scaleinvariant:vor_cells_distances}})
\begin{equation}
\label{eq:scaleinvariant:dS}
d_s(L,L')=\ln\{ \max\{ scale(L,L'),scale(L',L) \} \}
\end{equation}
The logarithm above has the base e. Any other base changes $d_s$ only by a constant factor. The scale-invariant distance can help to recognize similar crystal structures based on scaled motifs, for example when we 'extend' arms of the T2 molecule by adding benzene rings. Both new distances $d_H$ and $d_s$ are independent of the group symmetry of crystal structures. \\
\begin{thm}
\label{thm:scaleinvariant:satisfying_axioms}
(proved in Appendix). The scale-invariant distance $d_s$ is independent of a lattice representation, satisfies the axioms \ref{axioms:identity}, \ref{axioms:symmetry}, \ref{axioms:triangle_ineq} and both conditions \ref{axioms:continuity}, \ref{axioms:scaling}.
\end{thm}

\section{Computations of new distances for lattices of simulated T2 crystal structures}
\label{sec:newdistances_computation}
\textbf{\subsection{A fast algorithm to approximate the new distances $d_H$ and $d_s$ on any lattices}}
\label{sec:newdistances_computation:approx_dh_ds}

\begin{thm}
\label{thm:newdistances_computation:lineartime}
(proved in Appendix). For any polyhedra $P,P' \subset \R^n$ symmetric with respect to 0,
$$ \begin{gathered}
offset(P,P')=min\{ r>0:P \subset N(P';r)\} \: and\\ scale(P,P')=min\{ s>0: P \subset sP' \}
\end{gathered} $$
can be computed in a linear time with respect to the numbers of vertices and faces of \textit{P,P'}.
\end{thm}
The Voronoi cell of any lattice in $\R^3$ has at most 24 vertices and 14 faces, see \textbf{Figure \ref{fig:hex_sq_cub_bcc_fcc_vor}}. Hence Theorem 5 guarantees a constant upper bound not depending on lattices \textit{L,L'}. The standard way to optimize over all rotations $R \in SO(3)$ is to uniformly sample all rotations by using these parameters: a unit length vector $v \in \R^3$ and an angle $\theta \in [0,360\degree)$ of the rotation around \textit{v}. Take \textit{v} in the upper hemisphere (to avoid opposite vectors giving the same axis) as
\begin{equation}
\label{eq:newdistances_computation:rotationaxis}
v=(\sqrt{1-z^2} \cos{\mu},\sqrt{1-z^2} \sin{\mu}, z)
\end{equation}
If the height parameter $z \in (0,1)$ has n samples, then the angles $\mu,\theta$ have $[2 \pi n]$ samples. In total, about $4 \pi^2n^3$ rotations from \textit{SO(3)} will be sampled. The experiments below use n = 3, hence more than 1000 sampled rotations. Any vector $u \in \R^3$ is rotated by Rodrigues’ formula
\begin{equation}
\label{eq:newdistances_computation:rodriguez}
u \rightarrow u\:\cos{\theta}+(v \times u)\:\sin{\theta}+v(v \times u)(1-\cos{\theta}).
\end{equation}

\begin{figure}[h]
	\centering%
	\begin{subfigure}[b]{0.20\textwidth}
		\centering% <-- added
		\includegraphics[width=\textwidth, keepaspectratio]{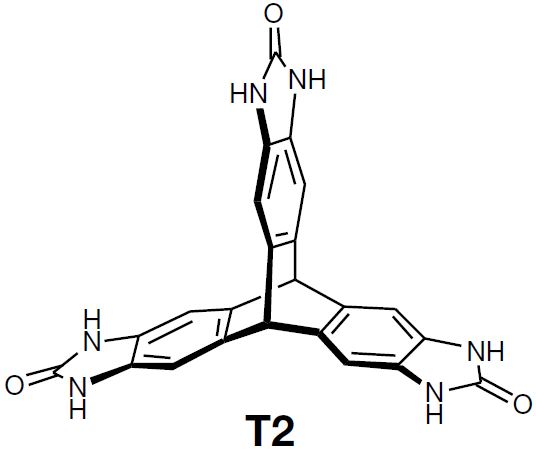}
		\label{fig:newdistances_computation:t2chemdiag}
	\end{subfigure}%
	\begin{subfigure}[b]{0.15\textwidth}
		\centering % <-- added
		\includegraphics[width=\textwidth, keepaspectratio]{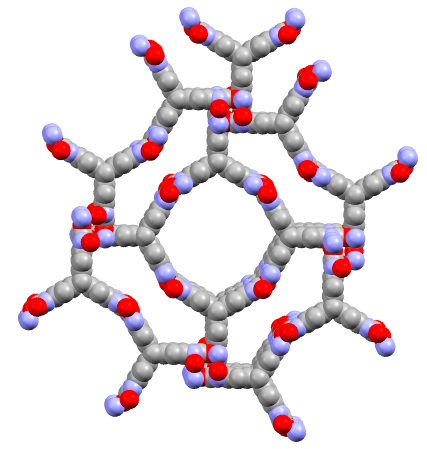}
		\label{fig:newdistances_computation:t2alpha}
	\end{subfigure}
	\begin{subfigure}[b]{0.08\textwidth}
		\centering % <-- added
		\includegraphics[width=\textwidth, keepaspectratio]{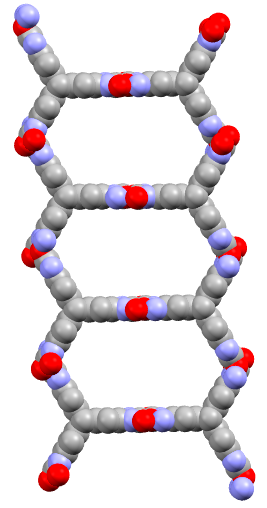}
		\label{fig:newdistances_computation:t2beta}
	\end{subfigure}
	\begin{subfigure}[b]{0.12\textwidth}
		\centering % <-- added
		\includegraphics[width=\textwidth, keepaspectratio]{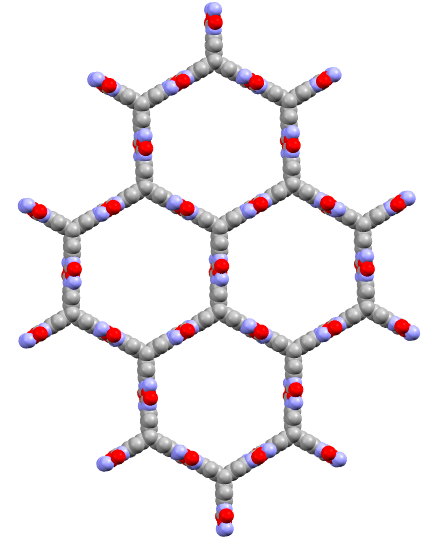}
		\label{fig:newdistances_computation:t2gamma}
	\end{subfigure}
	\begin{subfigure}[b]{0.18\textwidth}
		\centering % <-- added
		\includegraphics[width=\textwidth, keepaspectratio]{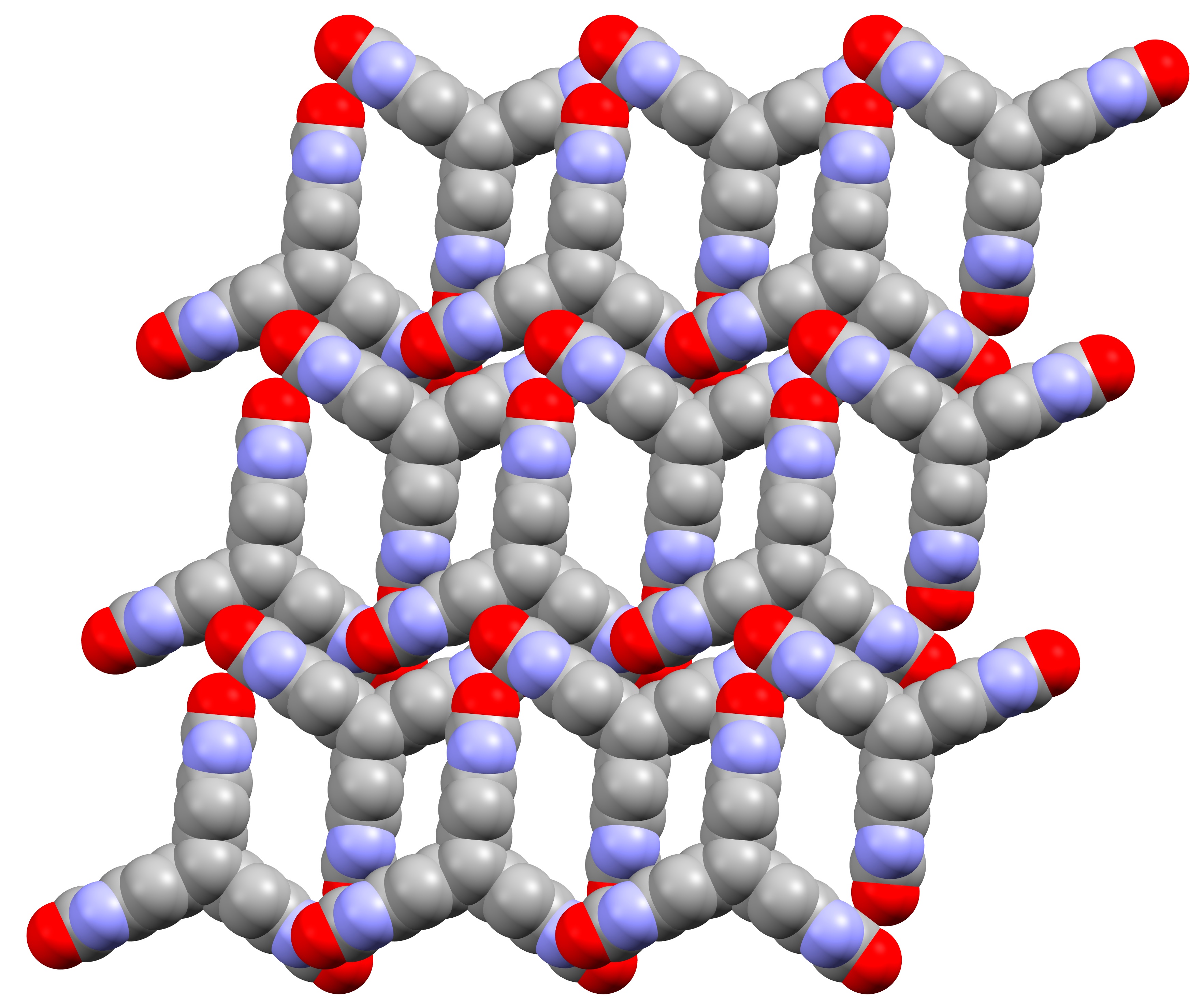}
		\label{fig:newdistances_computation:t2delta}
	\end{subfigure}
	\begin{subfigure}[b]{0.20\textwidth}
		\centering % <-- added
		\includegraphics[width=\textwidth, keepaspectratio]{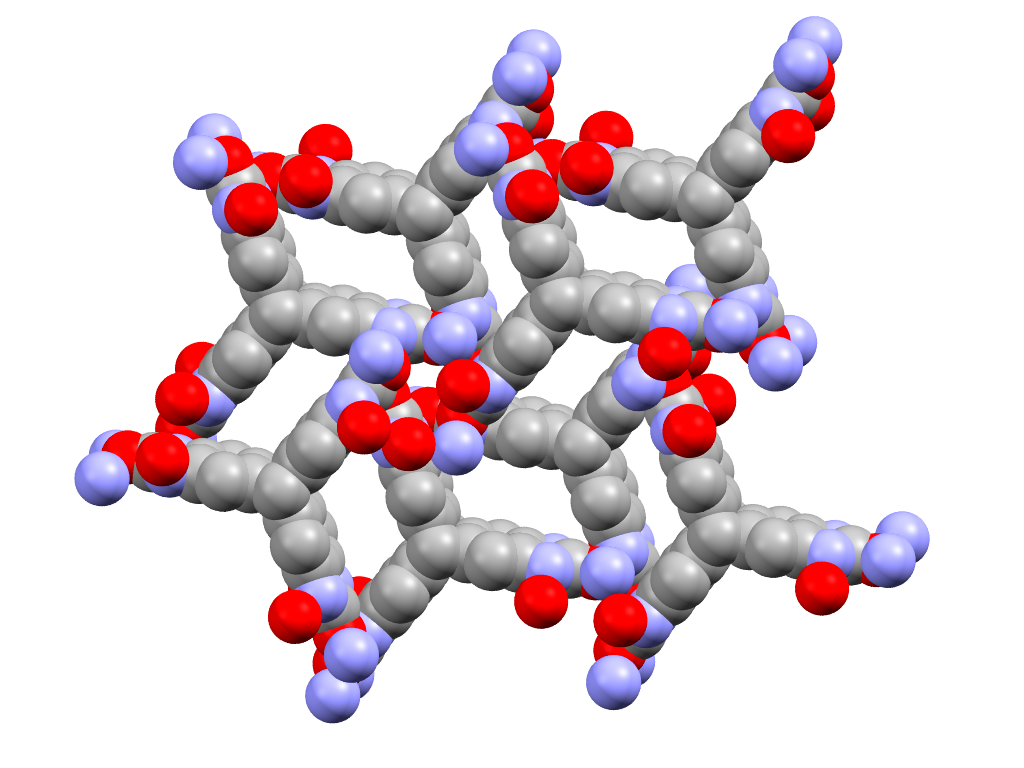}
		\label{fig:newdistances_computation:t2epsilon}
	\end{subfigure}
	%\captionsetup{justification=centering}
	\caption{(T2 molecule: triptycenetrisbenzimidazolon; crystals T2$\alpha$, T2$\beta$, T2$\gamma$, T2$\delta$, T2$\epsilon$. \cite{biblio:linjiangandy})}
	\label{fig:newdistances_computation:realcrystals}	
\end{figure}

\begin{figure}[h]
	\centering%
	\begin{subfigure}[b]{0.35\textwidth}
		\centering% <-- added
		\includegraphics[width=\textwidth, keepaspectratio]{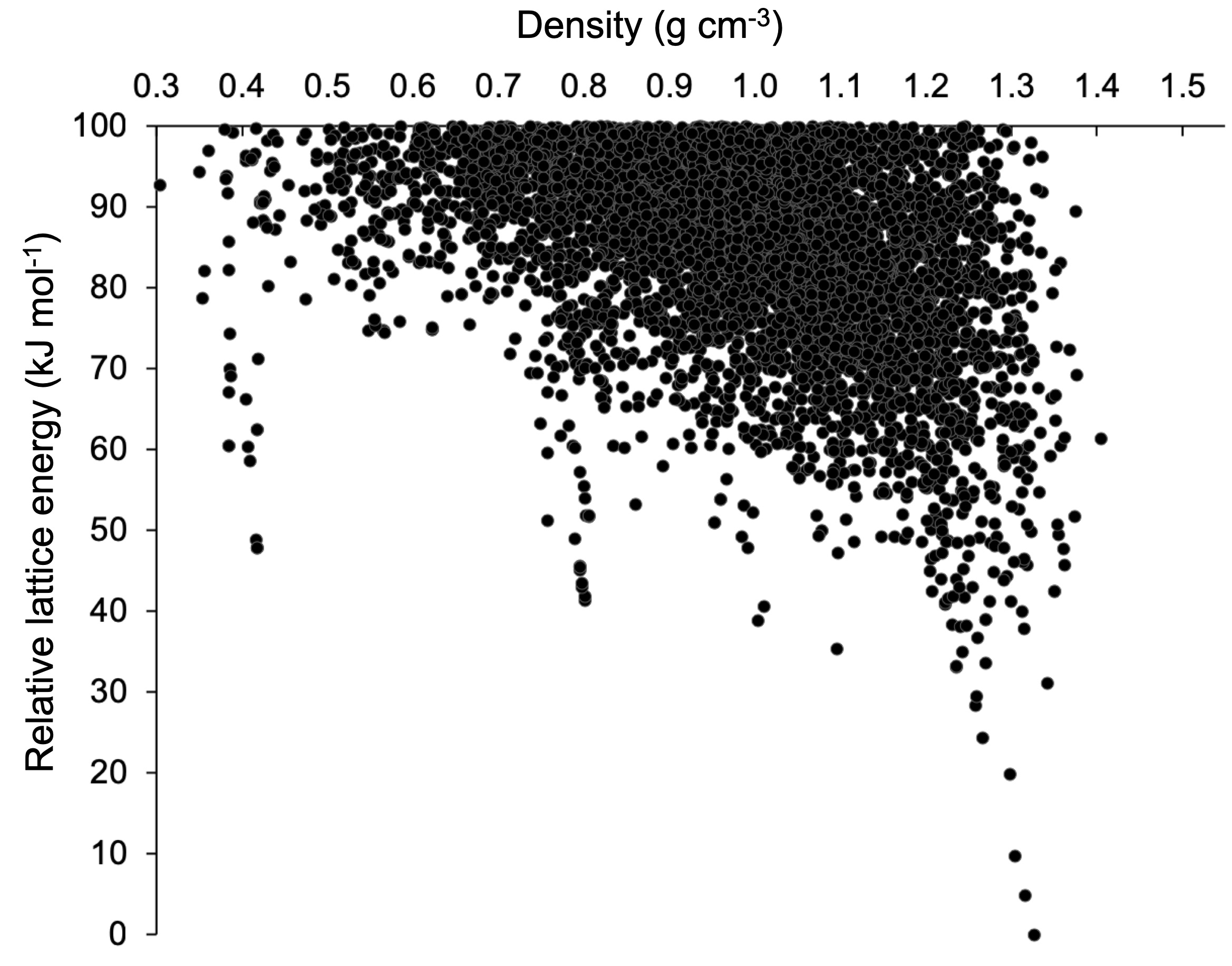}
		\label{fig:newdistances_computation:csp_land}
	\end{subfigure}%
	\begin{subfigure}[b]{0.3\textwidth}
		\centering % <-- added
		\includegraphics[width=\textwidth, keepaspectratio]{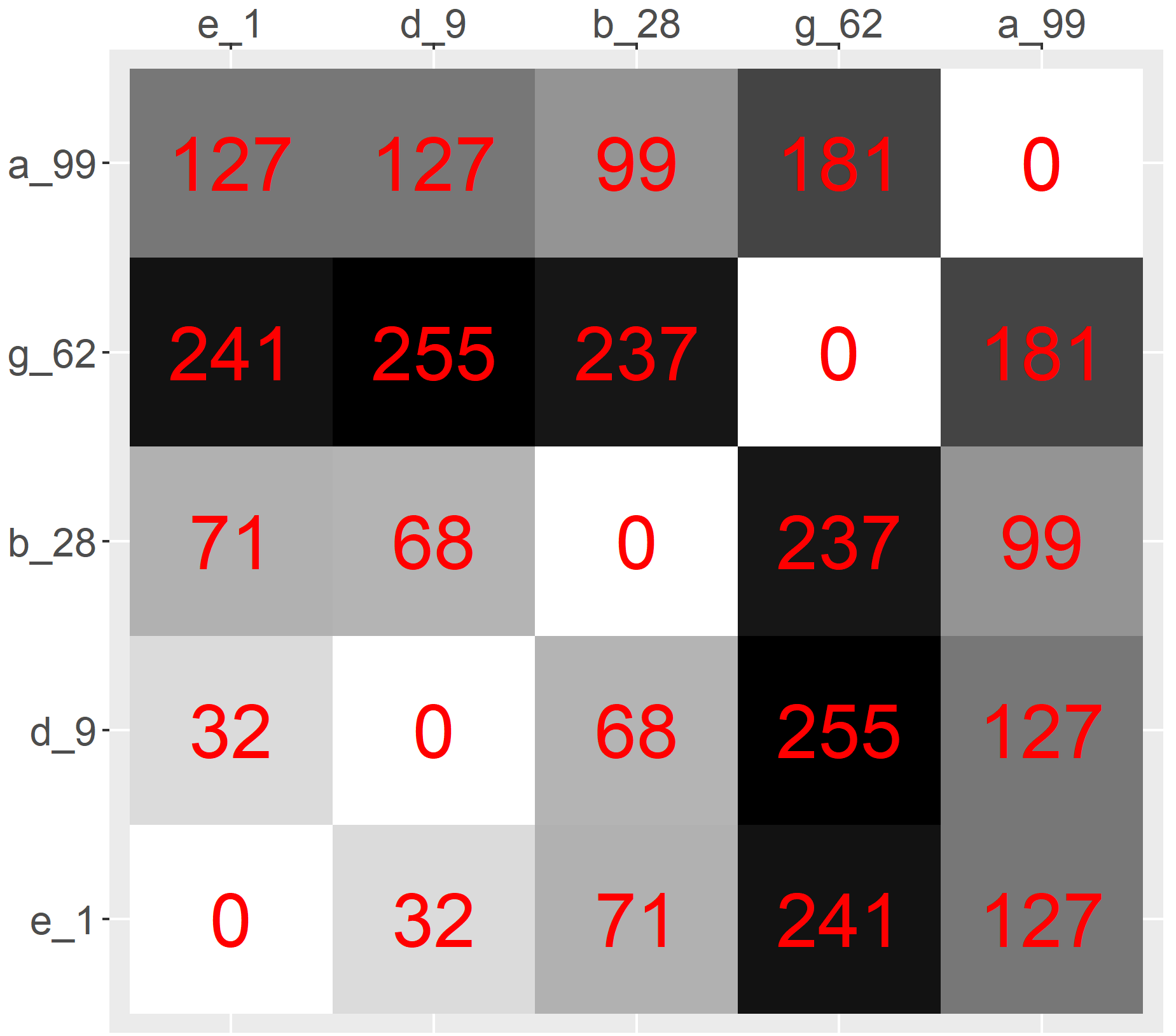}
		\label{fig:newdistances_computation:ext_hausdorff_realcrystals}
	\end{subfigure}
	\begin{subfigure}[b]{0.3\textwidth}
		\centering % <-- added
		\includegraphics[width=\textwidth, keepaspectratio]{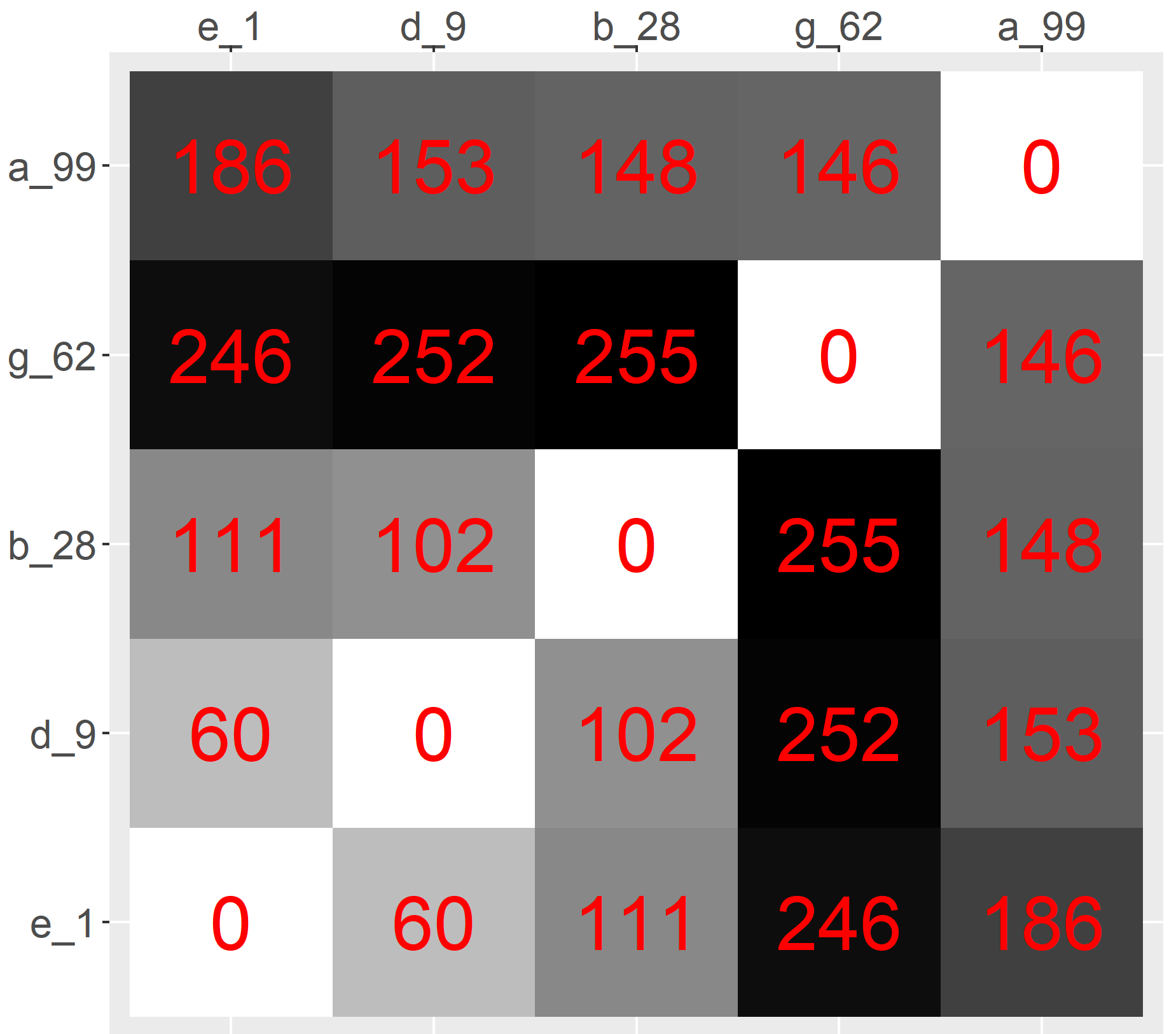}
		\label{fig:newdistances_computation:scaleinvariant_realcrystals}
	\end{subfigure}
	%\captionsetup{justification=centering}
	\caption{(Left: energy-vs-density plot of simulated structures.\cite{biblio:linjiangandy} Middle: extended Hausdorff distances $d_H$, Right: scale-invariant distances $d_s$ for real structures T2$\alpha(a\_99)$, T2$\beta(b\_28)$, T2$\gamma(g\_62)$, T2$\delta(d\_9)$, T2$\epsilon(e\_1)$. All values are scaled to [0,255] and rounded to integers.)}
	\label{fig:newdistances_computation:csp_dh_ds_reals_heatmaps}	
\end{figure}

\textbf{\subsection{Simulated and synthesized organic crystal structures based on the T2 molecule}}
\label{sec:newdistances_computation:t2crystals}
The Nature paper by Pulido et al. \cite{biblio:linjiangandy} has demonstrated that functional organic materials can be discovered by simulating crystals build on a molecule with a desired function via costly optimizations of the energy and target properties such as gas adsorption. Only 5 crystals in \textbf{Figure \ref{fig:newdistances_computation:realcrystals}} were synthesized, though predictions of properties were run for all crystals. The key bottleneck in this approach is the time-consuming prediction for nearly identical simulated crystals. For example, producing the CSP landscape energy-vs-density for 5688 crystals based on the T2 molecule has taken many weeks of the supercomputer time, see \textbf{Figure \ref{fig:newdistances_computation:csp_dh_ds_reals_heatmaps}}. The standard continuous similarity measures such as the energy and density are not enough to reliably quantify differences between crystals, because the same or almost identical energy and density cannot guarantee geometric similarity. The experiments on the T2 dataset of 5688 simulated crystal structures have found numerous pairs of crystals, e.g. with IDs (41,47), (68,71), (63,73), (71,83), (71,93), which have energy differences within $3 \frac{KJ}{mol}$ and also density differences within $0.01 \frac{g}{cm^{3}}$. However, these crystal structures have extended Hausdorff distances $d_H \geq 15$ Angstroms and scale-invariant distances $d_s \geq 1.1$, i.e. with scale factors more than $e^{1.1} \approx 3$. Crystal lattices 41 and 47 have unit cell angles close to 90\textdegree, but very different unit cell sides: (53.3, 23.7, 7.3), (15.4, 12.9, 16.5). These differences in their structure were found only now by using the new distances, not by the energy and density. Mercury has managed to match only 1 of attempted 50 T2 molecules in these structures, so the found deviation of positions is 0, because both crystals consist of the same T2 molecules.

\begin{figure}[h]
	\centering%
	\begin{subfigure}[b]{0.5\textwidth}
		\centering% <-- added
		\includegraphics[width=\textwidth, keepaspectratio]{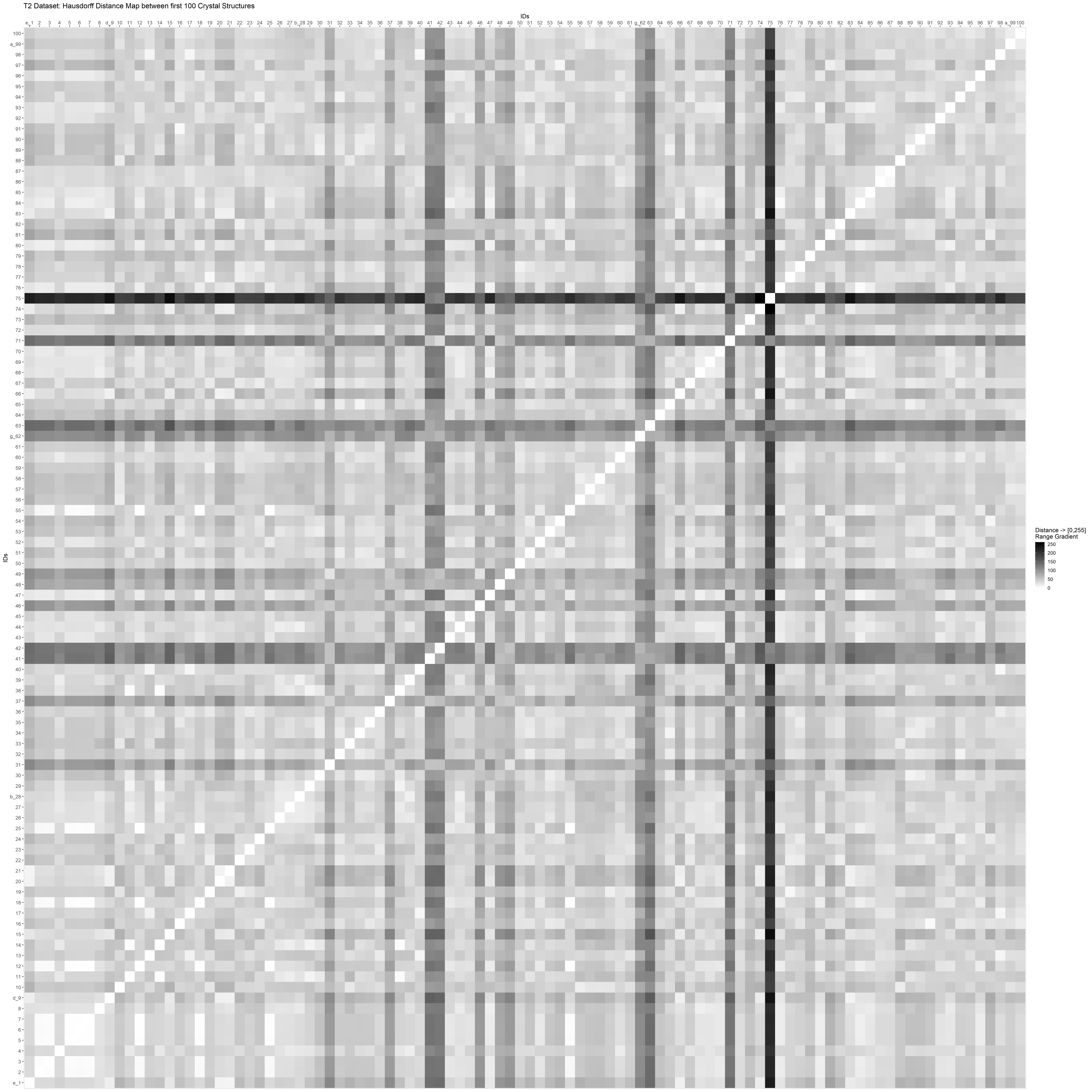}
		\label{fig:newdistances_computation:ext_dH_100x100}
	\end{subfigure}%
	\begin{subfigure}[b]{0.5\textwidth}
		\centering % <-- added
		\includegraphics[width=\textwidth, keepaspectratio]{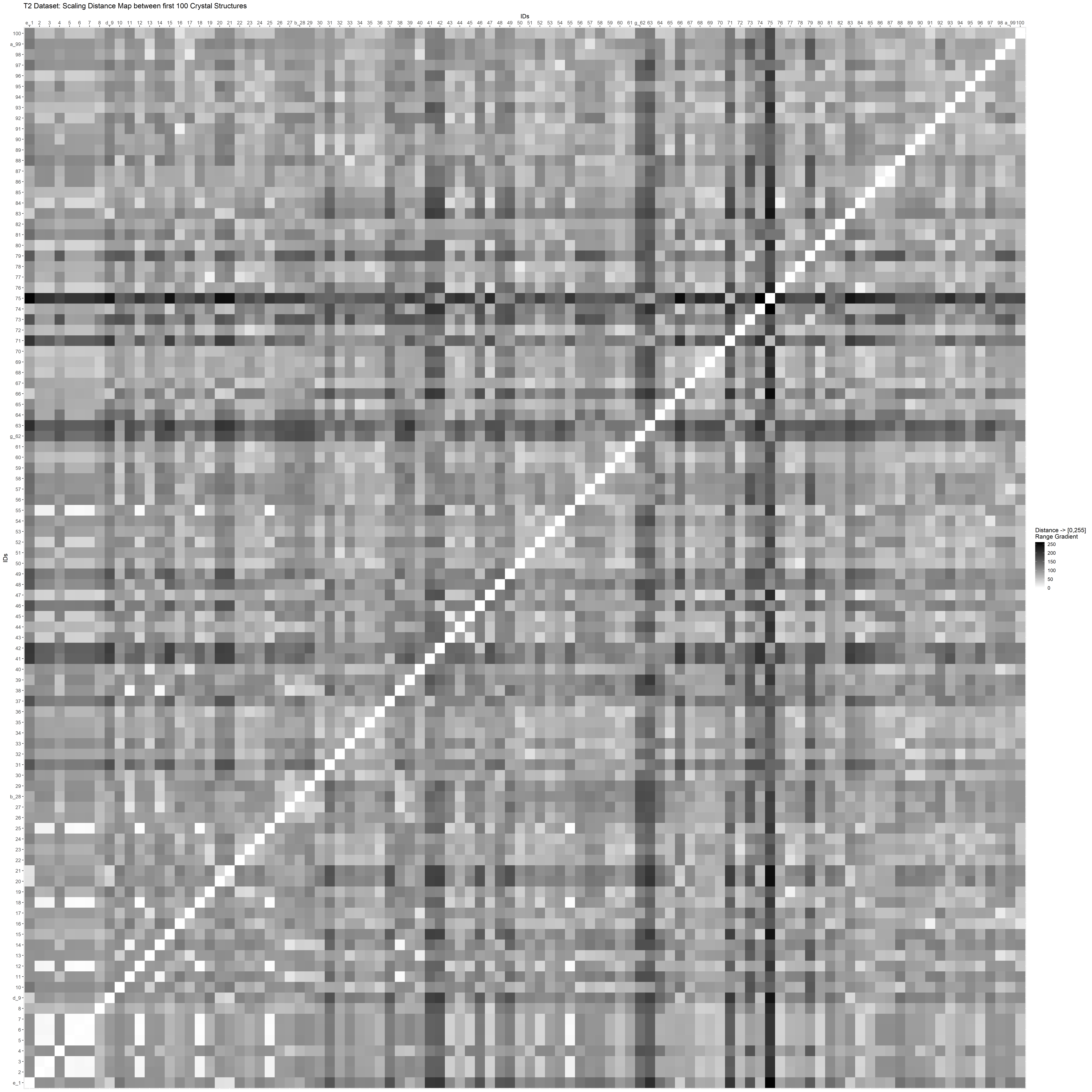}
		\label{fig:newdistances_computation:dS_100x100}
	\end{subfigure}	
	%\captionsetup{justification=centering}
	\caption{(Left: extended Hausdorff distances $d_H$, Right: scale-invariant distances $d_s$ for first 100 of 5688 simulated crystal structures reported in the Nature paper. \cite{biblio:linjiangandy} All values are linearly scaled to [0,255] and displayed in a grayscale heatmap similarly to Figure 6.)}
	\label{fig:newdistances_computation:extdH_dS_100x100}	
\end{figure}

\textbf{\subsection{Similarity matrices (grayscale maps) of the new distances on the T2 crystal lattices}}
\label{sec:newdistances_computation:similarity_matrices}
For each of the distances $d_H$ and $d_s$, all values (from a minimum to a maximum) were linearly scaled to the range [0,255] to visualize the distance matrix in grayscale. Figure \ref{fig:newdistances_computation:csp_dh_ds_reals_heatmaps} shows the distances (rounded to integers) between 5 real T2 crystal structures. \textbf{Figure \ref{fig:newdistances_computation:extdH_dS_100x100}} shows the larger experiment on the first 100 crystals from the T2 dataset. Both distances $d_H$ and $d_s$ were computed for all 4950 unordered pairs of these 100 crystals, which took in total 6 hours on a modest laptop. The variability of intensities in both heatmaps justifies that $d_H$ and $d_s$ take many values not restricted to a small discrete set. The full 100 $\times$ 100 matrices for $d_H$ and $d_s$ are in the supplementary materials. The scale-invariant distance $d_s$ shows more variability of colors in the heatmap in \textbf{Figure \ref{fig:newdistances_computation:extdH_dS_100x100}}, which confirms the usefulness of scale condition \ref{axioms:scaling}, which led to the new scale-invariant distance $d_s$. The C++ code “Lattice Distances” for the new distances $d_H$ and $d_s$ will be available soon.

\bigskip\bigskip
\section{Conclusions: contributions to the state-of-the-art justified by proofs and experiments}
\label{sec:conclusions}
\begin{itemize}
\item The new (extended Hausdorff $d_H$ and scale-invariant $d_s$) distances are defined for equivalence classes of arbitrary lattices considered up to any rigid motions in $\R^n$, hence are independent of choice of unit cells or coordinates in crystal representations.
\item Both distances satisfy the metric axioms and the continuity under perturbations proved in Theorems \ref{thm:ext_hausdorff:satisfying_axioms} and \ref{thm:scaleinvariant:satisfying_axioms}, which allows one to quantify similarities in a continuous way. Such a quantification is the important step to produce continuous hierarchies of crystal structures and visualize patterns of clusters changing for a varied distance threshold.
\item Experiments in section \ref{sec:newdistances_computation} on real and simulated T2 crystal structures \cite{biblio:linjiangandy} show that $d_H$ and $d_s$ better distinguish crystal lattices that have almost identical energy and density.
\end{itemize}

\noindent \textbf{Supporting Information}: all proofs, distance matrices and images are in the attached zip.\\
\begin{flushright}
	Received: ((will be filled in by the editorial staff))\\
	Revised: ((will be filled in by the editorial staff))\\
	Published online: ((will be filled in by the editorial staff))
\end{flushright}

%\paragraph{Paragraph headings} Use paragraph headings as needed.

%\begin{figure}
%  \includegraphics{images/bcc.png}
%\label{fig:_1}       % Give a unique label
%\end{figure}
%\begin{figure*}
%  \includegraphics[width=0.75\textwidth]{images/bcc.png}
%\label{fig:10}       % Give a unique label
%\end{figure*}
%\begin{table}
%\begin{tabular}{lll}
%\hline\noalign{\smallskip}
%\end{tabular}
%\end{table}

\begin{acknowledgements}
We thank the reviewers for helpful advice. The work is supported by the EPSRC grant “Application-driven Topological Data Analysis” (2018-2023), EP/R018472/1.
\end{acknowledgements}



\pagebreak

\begin{figure}[h]
	\centering% <-- added
	\captionsetup{justification=centering}
	\frame{\includegraphics[width=3cm,keepaspectratio]{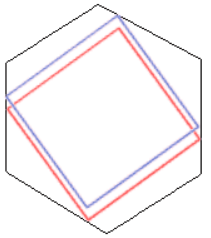}}
	\caption{(Let a centrally symmetric polyhedron $P'$ cover a centrally symmetric polyhedron $P$. Then the symmetric image of $P$ with respect to the center of $P'$ is also covered by $P'$.)}
	\label{fig:s1}
\end{figure}

\noindent \textbf{APPENDIX: Voronoi-based similarity distances between arbitrary crystal lattices}\\
Marco Mosca, Vitaliy Kurlin*\\
\\
\textbf{Proof of Lemma \ref{lem:ext_hausdorff:centr_sym_polyhedron_and_offset_param}}. Assume by contradiction that $\min\{r: T_v(P) \subset N(P';r)\}$ is minimized for a vector \textit{v} that differs from $c(P')-c(P)$, see \textbf{Figure \ref{fig:s1}}. Without loss of generality one can assume that $v=0$ and $r=0$, i.e. one needs to prove that if $P \subset P'$, then this inclusion is preserved when center $c(P)$ is shifted to $c(P')$. Under the symmetry (inversion) $S$ with respect to $c(P')$ the polyhedron \textit{P'} remains at the same position and covers the symmetric image $S(P)$ of $P$. The polyhedra $P,S(P)$ are connected by the continuous motion moving the center $c(P)$ to its symmetric image under S through the center $c(P')$. All intermediate images of $P$ remain covered by $P'$ due to the convexity of $P'$. Indeed, any two points belong to $P'$ together with the line segment connecting them. Hence the polyhedron $P$ shifted by the vector $c(P')-c(P)$ is also covered by $P'$. Symmetrically, if one fits a translational image of $P'$ into a minimal offset of $P$, then an optimal translation should make the centers of $P,P'$ identical.

\noindent \textbf{Proof of Theorem \ref{thm:ext_hausdorff:satisfying_axioms}}. The extended Hausdorff distance $d_H$ between lattices is based on the Voronoi cells, which are defined in terms of distances to lattice nodes, hence are independent of a linear bases of a lattice. By formula (\ref{eq:ext_hausdorff:offset}) $d_H(L,L')$ is always not negative and equals 0 only when there is a rotation $R$ such that $d_H(R(V(L)), V(L'))=0$, hence the Voronoi cells $V(L), V(L')$ become identical under the rotation $R$, so the lattices $L,L'$ are equivalent, which proves axiom \ref{axioms:identity}. Axiom \ref{axioms:symmetry} follows from formula (\ref{eq:ext_hausdorff:dH}) taking the maximum of two offsets when $L,L'$ are swapped. To check axiom \ref{axioms:triangle_ineq}, without loss of generality, one can assume that maxima in formula (\ref{eq:ext_hausdorff:dH}) are attained on first offsets. Fix optimal rotations $R,R'$ so that 
$$
\begin{gathered}
d_H(L,L')=offset(R(V(L)),V(L')) \: and\\ d_H(L',L'')=offset(R'(V(L')),V(L'')) 
\end{gathered}
$$
By (\ref{eq:hausdorff}), (\ref{eq:ext_hausdorff:offset}), (\ref{eq:ext_hausdorff:dH}) the first Hausdorff distance $d_H(L,L')$ above has a minimum value (say, \textit{r}) when
\begin{equation}
\label{eq:optrot_R_in_NL'}
\begin{gathered}
R(V(L)) \subset N(V(L');r) \: and \\
V(L') \subset N(R(V(L));r) \: or \: R^{-1}(V(L')) \subset N(V(L);r)
\end{gathered}
\end{equation}
Similarly, the second Hausdorff distance $d_H(L',L'')$ has a minimum value (say, \textit{r'}) when 
\begin{equation}
\label{eq:optrot_R'_in_NL''}
\begin{gathered}
R'(V(L')) \subset N(V(L'');r'), \\
V(L'') \subset N(R'(V(L'));r') \: or \: (R')^{-1}(V(L'')) \subset N(V(L');r')
\end{gathered}
\end{equation}
The composition of rotations $R'R$ (applied from right to left) rotates the Voronoi cell $V(L)$ to the position $R'R(V(L))$. The first inclusions from (\ref{eq:optrot_R_in_NL'}), (\ref{eq:optrot_R'_in_NL''}) above imply the inclusion below:
\begin{equation}
\label{eq:optrot_R'R_in_NL''}
\begin{gathered}
R'R(V(L)) \subset \\
\subset R'(N(V(L');r))=N(R'(V(L'));r) \subset\\
\subset N(N(V(L'');r');r)=N(V(L'');r+r')
\end{gathered}
\end{equation}
Similarly, the opposite composition of rotations $R^{-1}(R')^{-1}$ (applied from right to left) rotates the Voronoi cell $V(L'')$ to $R^{-1}(R')^{-1}(V(L''))$. The last inclusions from (\ref{eq:optrot_R_in_NL'}), (\ref{eq:optrot_R'_in_NL''}) imply that 
\begin{equation}
\label{eq:optrot_R-1R'-1_in_NL}
\begin{gathered}
R^{-1}(R')^{-1}(V(L'')) \subset \\ 
\subset R^{-1}(N(V(L');r'))=N(R^{-1}(V(L'));r') \subset\\
\subset N(N(V(L);r);r')=N(V(L);r+r')
\end{gathered}
\end{equation}
Inclusions (\ref{eq:optrot_R'R_in_NL''}), (\ref{eq:optrot_R-1R'-1_in_NL}) mean that the extended Hausdorff distance $d_H(L,L'')$ has the upper bound $r+r'$ attained for the rotations $R'R$ and $R^{-1}(R')^{-1}$. The minimum over all rotations can be even smaller, hence the triangle inequality $d_H(L,L'')\leq r+r'$ in axiom \ref{axioms:triangle_ineq} holds. Continuity condition \ref{axioms:continuity} follows from the stability of Voronoi cells in subsection \ref{sec:newdistances:vcell_stability}.\\
\\
\textbf{Proof of Lemma \ref{lem:scaleinvariant:min_s_centers_superimposed}} is similar to the proof of Lemma \ref{lem:ext_hausdorff:centr_sym_polyhedron_and_offset_param} with \textit{r}-offsets $N(P;r)$ replaced by scaled polyhedra $s\:P \subset \R^n$, because all other inclusion and convexity arguments remain valid.\\
\\
\textbf{Proof of Theorem \ref{thm:scaleinvariant:satisfying_axioms}}. Similarly to the proof of Theorem \ref{thm:ext_hausdorff:satisfying_axioms}, the scale-invariant distance $d_s$ between lattices is based on the Voronoi cells and is independent of lattice representations.
To check that $d_s(L,L') \geq 0$, let $R,R'$ be optimal rotations that minimize the factors $s=scale(L,L')$ and $s'=scale(L',L)$, respectively. Formula (\ref{eq:scaleinvariant:scale}) implies that $$R'R(V(L)) \subset R'(s\:V(L'))=s\:R'(V(L')) \subset s\:s'\:V(L)$$ 
Since the volumes of $R'R(V(L))$ and $V(L)$ are equal, the inclusion $R'R(V(L)) \subset s\:s'\:V(L)$ implies that $ss'\geq1$, hence $\ln( \max \{ s,s' \} )\geq0$. The equality is possible only if both $s=s'=1$, which means that $V(L),V(L')$ are obtained from each by a rotation, hence the lattices $L,L'$ are equivalent, so axiom \ref{axioms:identity} is proved. Axiom \ref{axioms:symmetry} follows from symmetric formula (\ref{eq:scaleinvariant:dS}). 
To check axiom \ref{axioms:triangle_ineq}, one can assume without loss of generality that the minimal scales are attained on the first scales among two in formula (\ref{eq:scaleinvariant:dS}). Fix optimal rotations $R,R'$ such that $scale(L,L')$ and $scale(L',L'')$ take minimum values (say, s and s', respectively) when $$R(V(L)) \subset s\:V(L') \: and \: R'(V(L')) \subset s'\:V(L'')$$. Then $$R'R(V(L)) \subset R'(s\:V(L'))=s\:R'(V(L')) \subset s\:s'\:V(L'')$$. Hence $scale(L,L'') \leq s\:s'$, because an optimal rotation from $V(L)$ to $V(L'')$ may have a smaller scale than achieved by $R'R$. The symmetric $scale(L'',L)$ has a similar upper bound from optimal rotations or $scale(L'',L)$ and $scale(L',L)$. The triangle inequality follows after taking the logarithm of both sides:
\begin{align*}
\max\{ scale&(L,L''),scale(L'',L)\} \leq\\ 
&\leq \max\{ scale(L,L'),scale(L',L) \} \: \max\{ scale(L',L''),scale(L'',L') \}
\end{align*}

To prove continuity condition \ref{axioms:continuity} let r(L) be the distance from the origin $0 \in L$ to the boundary of the Voronoi cell $V(L)$. The geometric stability of Voronoi cells in subsection 4.2 guarantees that the Voronoi cell $V(L')$ of a perturbed lattice \textit{L'} is in the \textit{r}-offset $N(V(L);r)$ of $V(L)$ for a small $r>0$. Since centrally symmetric Voronoi cells of \textit{L,L'} are compared below, one can assume that their centers coincide with the origin in $\R^{n}$. For any $p \in N(V(L);r)$, let $R(0,p)$ be the straight ray emanating from 0 and passing through p. Let \textit{q} be the intersection of $R(0,p)$ with the boundary of $V(L')$. The ratio $\frac{d(p,q)}{d(0,q)}$ is at most $\frac{r}{r(L)}$, hence $$d(p,0)=d(p,q)+d(0,q) \leq d(0,q)(1+\frac{r}{r(L)})$$
Then $$V(L') \subset N(V(L);r) \subset (1+\frac{r}{r(L)})V(L) \: and \: scale(L',L) \leq (1+\frac{r}{r(L)})$$
Swapping \textit{L,L'}, we get the upper bound for the scale-invariant distance $$d_s(L,L') \leq \ln(1+\frac{r}{\min\{r(L),r(L')\}})$$, which means that \textit{L'} remains close to \textit{L}. The scale-invariant in \ref{axioms:scaling} holds by formula (\ref{eq:scaleinvariant:scale}), because the inclusion $R(V(L)) \subset s\:V(L')$ remains unchanged then both lattices \textit{L,L'} are simultaneously scaled by the same factor.\\
\\
\textbf{Proof of Theorem \ref{thm:newdistances_computation:lineartime}}. The minimum offset $$offset(P,P')=\min\{ r>0:P \subset N(P';r) \}$$ is computed by starting from $r=0$ and updating \textit{r} for every vertex \textit{v} of \textit{P} as follows. Find the intersection of the line segment $[0,v]$ from the origin 0 to $v$ with a face \textit{F} of \textit{P'}. If there is such an intersection, then \textit{r} increases to the distance $d(v,F)$. Similarly, $$scale(P,P')=\min\{ s>0:P \subset s\:P'\}$$ is computed by finding the minimum scale \textit{s} that is enough to keep every vertex \textit{v} of \textit{P} inside \textit{P'}. Find the intersection of the ray $R(0,v)$ going from the origin 0 and passing via \textit{v} with a face \textit{F} of \textit{P'}, then \textit{s} increases to $\frac{d(0,v)}{d(0,R(0,v) \cap F)}$. In the worst case, intersecting lines through vertices of \textit{P} and flat faces of \textit{P'} requires a loop over all vertices and a loop over all faces. An upper bound for the asymptotic complexity is the product of the numbers of vertices and faces, which is a linear function in each number.
% end of file template.tex
\end{document}